\documentstyle[12pt]{article}
\newcommand{\be}{\begin{equation}}
\newcommand{\ee}{\end{equation}}
\newcommand{\ba}{\begin{eqnarray}}
\newcommand{\ea}{\end{eqnarray}}
\newcommand{\baa}{\begin{eqnarray*}}
\newcommand{\eaa}{\end{eqnarray*}}
\newcommand{\bb}{\begin{thebibliography}}
\newcommand{\eb}{\end{thebibliography}}
\newcommand{\ci}[1]{\cite{#1}}
\newcommand{\bi}[1]{\bibitem{#1}}
\newcommand{\lab}[1]{\label{#1}}
\title{
{\bf The energy dependence of the Pomeron and the eikonalized Pomeron }}
\author{\bf S.V.Goloskokov, S.P.Kuleshov, O.V.Selyugin\\
Bogoliubov Laboratory of Theoretical Physics \\
Joint Institute for Nuclear Research, Dubna }
\begin{document}
\phantom{.}
\vspace {1.cm}
\begin{center}
{\large { \bf The energy dependence of the Pomeron
                       and the eikonalized Pomeron }}
\vspace {.5cm}

{\bf S.V.Goloskokov, S.P.Kuleshov, {\underline {O.V.Selyugin}}   }\\
\vspace {.2cm}
Bogoliubov Laboratory of Theoretical Physics \\
Joint Institute for Nuclear Research, Dubna \\
\vspace {.2cm}
{\bf Abstract}
\end{center}
The one-to-one connection between the
eikonal phase and the ratio of the elastic and total cross section is
shown.
   Based on new experimental data of Collaboration CDF we analyzed
intercept and power of the logarithmic growth of the Born and
total Pomeron amplitude.

\vspace{1.5cm}

    Now we have a long and plenty discussion of the energy dependence
of the elastic and total cross section in the hadron - hadron
scattering \ci{blois,mart}.
Some models with the different hypothesis (see for example \ci{enk1}.
lead to
$$ \sigma_{tot} \sim ln s .$$
One of the recent analysis of the plenty experimental material has been made
in \ci{block1}. The conclusion was that this analysis gives
strong evidence of a $\log(s/s_0)$ dependence at current energies rather than
of $\log^{2}(s/s_0)$ and demonstrates that the odderon is not
necessary to explain experimental data.
But many models (for example \ci{nik1}) were based on the idea
of "supercritical" Pomeron
exchange, with $\varepsilon = \alpha(0) -1 > 0$, which after unitarization
results in the Froissart saturation of the total cross sections
$$ \sigma_{tot} \sim ln^{2} s. $$

     In \ci{levin} the overall analysis of experimental material was made
>from viewpoint of the soft and hard supercritical Pomeron where
the $\log^{2}(s/s_0)$ was obtained.
Good agreement with data on deep inelastic scattering
and photoproduction data is reached in the case when the non-perturbative
component of the Pomeron is governed by the "maximal" behaviour, i.e.
like $\ln^2(s)$ \ci{enk2}.

	One can make different remarks on this and others analyses
and on the values of the experimental material used. Essential
uncertainty in the values of $\sigma_{tot}$ was shown in \ci{sel1}.
Now we have the large discussion about of the value of $\sigma_{tot}$
at $\sqrt{s} = 1.8 TeV$. The last work of the CDF Collaboration \ci{CDF}
gives
$$ (1+\rho^2) \sigma_{tot}=62.64 \pm 0.95 (mb) \ \ at \ \ \sqrt{s}=546 GeV,$$
$$ (1+\rho^2) \sigma_{tot}=81.83 \pm 2.29 (mb) \ \ at \ \ \sqrt{s}=1.8 TeV,$$
and
$$ \delta(s_1)= \sigma_{elast} / \sigma_{tot} = 0.210 \pm .002 \ \
at  \ \ \sqrt{s}=546 GeV,$$
$$ \delta(s_2)= \sigma_{elast} / \sigma_{tot} = 0.246 \pm .004 \ \
at \ \ \sqrt{s}=1.8 TeV.$$

Let is compare  this value of $\sigma_{tot}$ at $ \sqrt{s}=1.8 TeV.$
with the previous value which was
equals to $72 mb$
     The lasts two relations have small errors
as a consequence of the cancellation of some errors.
As we will show further, the relation of these two values is more interesting
and helps us to obtain the intercept of the Born and eikonalized Pomeron.
Let us denote the ratio of these two last values by $\Delta(s_{12})$
$$  \Delta(s_{12})= \frac{\delta(s_1)}{\delta(s_2)}=
                  \frac{\sigma_{el}(s_1) \cdot \sigma_{tot}(s_2)}
			 {\sigma_{el}(s_2) \cdot \sigma_{tot}(s_1)}. $$
    Now we consider the ordinary representation for the Born term of the
scattering amplitude at $\sqrt{s} \geq 540 GeV$.
At such large energies, we can neglect the contribution of the non-leading
Regge terms, and in this analysis, for the simplicity, we neglect the real
part of the scattering amplitude
\be
T(s,t) = i h s^{\alpha(t)-1} e^{R^2_{0} \cdot t /2} \lab{tt}
\ee
with the linear trajectories $\alpha(t)=\alpha(0)+\alpha_{1} t$ and
$\alpha(0)=1 + \varepsilon$.
The differential elastic, total elastic and total  cross section have
the following forms:
$$  \frac{d\sigma}{dt}= \pi \mid T(s,t) \mid ; \ \ \
   \sigma_{el}= \int_{-\infty}^{0} \frac{d\sigma}{dt} dt; \ \ \
   \sigma_{tot}= 4\pi Im T(s,0). $$

       For the amplitude (\ref{tt}) we obtain
\be
      \sigma_{el} = \pi  h^2 \int_{-\infty}^{0} s^{2 \cdot (\alpha(t)-1)}
                         e^{R^2_0 \cdot t} dt
   \ \	 = \ \  2\pi \frac{ h^2 s^{2 \varepsilon} }{R^2}  \lab{elt}
\ee
and
\be
       \sigma_{tot}= 4\pi h s^{\varepsilon}
\ee
where $ R^2 = R^2_{0} \cdot (1 + \alpha \cdot ln s) $.

   Hence, the relation $\sigma_{el}/\sigma_{tot}$ is
\be
   \frac{\sigma_{el}}{\sigma_{tot}}= \delta(s) =
    \frac{h}{2 R^2} \cdot s^{\varepsilon}.
\ee
 Using the value of $\Delta(s_{12})$ we can find the intercept
of the Pomeron
\be
    \varepsilon= \frac{ ln(\Delta (s_{12}) )}{ln(s_1/s_2)} +
      ln[\frac{1+\alpha ln(s_1)}{1+\alpha ln(s_2)}] / ln(s_1/s_2)  \lab{et}
\ee
$$   =\varepsilon_{0}+\varepsilon_{1}.       $$

      If we  take the amplitude in the form
\be
T(s,t) = i h \cdot ln^{n}(s) \cdot e^{R^2_{0} \cdot t /2}, \lab{tb}
\ee
we can calculate the value of $n$.
\be
    n= \frac{ ln(\Delta (s_{12}) )}{ln(ln(s_1)/ln(s_2))} +
      ln[\frac{1+\alpha ln(s_1)}{1+\alpha ln(s_2)}]
                        / ln(ln(s_1)/ln(s_2)).    \lab{nt}
\ee
$$	 \ \ \ \      = n_0 + n_1                        $$

   Now let us consider the total eikonal amplitude
\be
  T(s,t)= i \  \int \rho d\rho J_{0}(\rho\Delta)(1-e^{i\chi(s,\rho}),
\ee
where
\be
 i \chi(s,\rho)= i \int \Delta J_{0}(\rho\Delta) T^{B}(s,-\Delta^2) d\Delta.
\ee

   For the eikonal phase we derive
\be
 i \chi(s,\rho)= - \frac{h s^{\varepsilon^{B}}}{R^2} e^{-\rho^2/(2R^2)} =
 -X \cdot e^{-\rho^2/(2R^2)} .
\ee
It can be shown that the total and
elastic cross section can be represented as
\be
 \sigma_{tot}=  R^2 \ \sum_{k=1}^{\infty}(-1)\frac{(-X)^k}{k!k},
\ee

\be
 \sigma_{el}=  R^2 \ \{ 2\sum_{k=1}^{\infty}(-1)\frac{(-X)^k}{k!k}
		       - \sum_{k=1}^{\infty}(-1)\frac{(-2X)^k}{k!k} \}.
\ee
Hence, its ratio depends only on $X$.
The calculation confirms this and, for the experimental
values we have the one-to-one correspondence with $X_i$.
$$ \delta(546)=.210 \leftrightarrow  X(546)=1.38$$
$$ \delta(1800)=.246 \leftrightarrow  X(1800)=1.862$$

Based on these values $X_i$, we can obtain the intercept $\varepsilon^{B}$
or a power of the logarithmic growth $n^{B}$.
They have nearly the same form as  (\ref{et}),(\ref{nt})
\be
    \varepsilon^{B}= \frac{ ln[X(s_1)/X(s_2)]}{ln(s_1/s_2)} +
      ln[\frac{1+\alpha ln(s_1)}{1+\alpha ln(s_2)}] / ln(s_1/s_2)  \lab{eb}
\ee
$$   \ \ \ \    = \varepsilon^{B}_{0} + \varepsilon_{1};   $$

\be
    n^{B}= \frac{ ln[X(s_1)/X(s_2)] }{ln(ln(s_1)/ln(s_2))} +
      ln[\frac{1+\alpha ln(s_1)}{1+\alpha ln(s_2)}] / ln(ln(s_1)/ln(s_2))
\lab{nb}
\ee
$$       =n^{B}_{0}+n_{1}.                                         $$

   Using the recent experimental data we can calculate
$$ \varepsilon_{0}=0.066; \ \ \ \    \varepsilon^{B}_{0}=0.126. $$
$$ n_{0}=0.913 \ \ \ \ n_{0}^{B}=1.73.$$

     The values of $\varepsilon_{1}$ and $n_{1}$ weakly depend
on the value of $\alpha$.
The values of the total cross sections
heavily depend on these values of  $\alpha$.

So, for $R^2=R^2_{0}(1+\alpha ln(s))$ we have \\

for  $\alpha=0.06$
\begin{tabbing}
$R^2_0=$ \hspace{1.5cm} \= 6.8 \hspace{.5cm}  \= 6.9\hspace{.5cm}
                        \= 7.0 \hspace{.5cm}  \= 7.1 \hspace{.5cm}
                                              \\
$\sigma_{tot}(546)=$ \> 59.5 \> 60.37 \> 61.24\> 62.12  \\
$\sigma_{tot}(1800)=$ \>79.44 \> 80.6\> 81.8 \>82.94  \\
\end{tabbing}
\medskip

for  $\alpha=0.05$
\begin{tabbing}
$R^2_0=$ \hspace{1.5cm} \= 7.3 \hspace{.5cm}  \= 7.4\hspace{.5cm}
                        \= 7.5 \hspace{.5cm}  \= 7.6 \hspace{.5cm}
                                              \= 7.7 \\
$\sigma_{tot}(546)=$ \> 59.3 \> 60.1 \> 60.91\> 61.72 \>62.54 \\
$\sigma_{tot}(1800)=$ \>78.55 \> 79.6\> 80.7 \>81.8 \>82.85 \\
\end{tabbing}
\medskip

    for  $\alpha=0.04$
\begin{tabbing}
$R^2_0=$   \hspace{1.5cm}  \= 7.8 \hspace{.5cm}     \= 8.0 \hspace{.5cm}   \=
8.2 \\
$\sigma_{tot}(546)=$                  \> 58.45   \> 59.9  \> 61.45 \\
$\sigma_{tot}(1800)=$                 \> 76.74   \> 78.7  \> 80.7
\end{tabbing}

  Let us  remember that the experimental values of $\sigma_{tot}$ are \\
$ (1+ \rho^2 ) \sigma_{tot} (546.) = 62.64 (mb) $ CDF \\
$ (1+\rho^2) \sigma_{tot} (546.) =63.5 (mb) $ UA4 \\
$ (1+\rho^2) \sigma_{tot} (1800) =81.83 (mb) $ CDF.\\

       It is clear from these Tables and Fig.1
that the value of  $\alpha$ cannot exceed $0.05$,
otherwise we will get the wide divergence from the experimental data
of the $\sigma_{tot}$. Moreover, we can conclude that the experimental
value of  $\sigma_{tot}$ at $\sqrt{s} = 1800 GeV$
cannot be less then $78 \div 80 (mb)$
or it will contradict
either the value of the ratio $\sigma_{el}/\sigma_{tot}$ or the value of
$\sigma_{tot}$ at $\sqrt{s} = 546 GeV$.

 The calculation gives for $\alpha=0.04$
$$ \varepsilon_{1}=0.0257 \ \ \ \  n_{1}=0.35 $$
and for $\alpha=0.05$
$$ \varepsilon_{1}=0.0296 \ \ \ \  n_{1}=0.407. $$
Hence, as the result we have for $\alpha=0.05$
$$ \varepsilon = 0.066+0.03 = 0.096; \ \ \ \
\varepsilon^{B}=0.126+0.03=0.156;$$
$$ n = 0.926+0.407=1.333; \ \ \ \  n^{B}=1.73+0.41=2.14. $$

   The situation does not change if we consider
$ R^2=R^2_{0}(1+\alpha \sqrt{ln(s)})$ (see \ci{levin}).
In this case, the compromising value
of  $\alpha = 0.4$ and for that value we find
$$ \sigma_{tot}(\sqrt{s} = 546 GeV) = 61.8 mb; \ \ \ \
    \sigma_{tot}(\sqrt{s} = 1800 GeV) = 80.72 mb $$.
and as a result we have practically the same values
$$ \varepsilon_{1} = 0.022  ;\ \ \ \ n_{1} = 0.3. $$

    So we can see that in the examined energy range the power of
the logarithmic growth of $\sigma_{tot}$ is larger than $1$
but sufficiently
smaller than $2$. It is clear that we have the $ln^2$
term in the total cross section but with
a small coefficient, and now we are very far from the asymptotic range.
The ratio $\sigma_{el}/\sigma_{tot} = .246$ says us the same.

    In our opinion, this calculation shows that the value of
$\varepsilon^{QCD} =0.15 \div 0.17$,
calculated in the framework of the QCD \ci{Braun}
 does not contradict the value of the
ordinary Pomeron intercept $ = 0.08$.
The $\varepsilon^{QCD}$ is to be  compared with the intercept of the
eikonalized Pomeron - $\varepsilon^{B}$ that is equal to $0.156$,
as it is evident from our analysis of experimental data.

      Our calculation was carried out
with the gaussian form of the scattering amplitude.
We think that our conclusion does not heavily depend on the
definite form of the diffraction peak and we hope to confirm this in
the neares future.
\vspace{0.5cm}

     {\it Acknowledgement.} {\hspace {0.5cm} The authors expresses his deep
gratitude to D.V.Shirkov, A.N. Sissakian and V.A. Meshcheryakov
for support in this work.

    This work was supported in part by the Russian Fond of
Fundamental Research, Grand $   94-02-04616$.

\vspace{1cm}

{\bf Figure Captions}

\phantom{.} \hspace{1cm} {\bf Fig.1} The dependence of
$\sigma_{tot}$ on $R_{0}$ for the
different values of $\alpha$.    \\
\phantom{.} \hspace{1cm} (the long-dash lines are for
$\sigma_{tot}(1800)=80.03 mb$ and
$\sigma_{tot}(546)=61.26 mb$  \\
\phantom{.} \hspace{1cm} the short-dash lines are for $R_{0}(i)$ such that
$\sigma_{tot}(s,R_{0})
=\sigma_{tot}(s)_{eksp.}$ )

\end{document}